\documentclass{article}
\usepackage{spconf, amsmath, amsthm, graphicx, soul, xcolor, amssymb, multirow, booktabs, subcaption, caption, array, adjustbox, float}
\usepackage[breaklinks=true, colorlinks, linkcolor = blue, urlcolor  = blue, citecolor = blue, anchorcolor = blue, bookmarks=true]{hyperref} 
\usepackage[ruled,vlined,linesnumbered]{algorithm2e}
\usepackage[switch]{lineno}

\def\L{{\cal L}}
\newcolumntype{P}[1]{>{\centering\arraybackslash}p{#1}}

\renewcommand{\L}{\mathcal{L}}

\newcommand{\norm}[1]{\left\lvert\left\lvert#1\right\rvert\right\rvert}
\newcommand{\la}{\left\langle}
\newcommand{\ra}{\right\rangle}

\title{SCP-GAN: Self-Correcting Discriminator Optimization for Training Consistency Preserving Metric GAN on Speech Enhancement Tasks}
\name{Vasily Zadorozhnyy$^{*,1}$\thanks{$^*$ Work performed while Vasily Zadorozhnyy was an intern at Microsoft.}, Qiang Ye$^{1}$, and Kazuhito Koishida$^{2}$}
\address{$^{1}$ Department of Mathematics, University of Kentucky, Lexington, USA\\
 $^{2}$ Applied Sciences Group, Microsoft Corporation, Redmond, USA\\
 $^{1}$\texttt{\{vasily.zadorozhnyy, qye3\}@uky.edu} \quad $^{2}$\texttt{kazukoi@microsoft.com}}

\begin{document}
%
\maketitle
\begin{abstract}
In recent years, Generative Adversarial Networks (GANs) have produced significantly improved results in speech enhancement (SE) tasks. They are difficult to train, however. In this work, we introduce several improvements to the GAN training schemes, which can be applied to most GAN-based SE models. We propose using consistency loss functions, which target the inconsistency in time and time-frequency domains caused by Fourier and Inverse Fourier Transforms. We also present self-correcting optimization for training a GAN discriminator on SE tasks, which helps avoid ``harmful" training directions for parts of the discriminator loss function. We have tested our proposed methods on several state-of-the-art GAN-based SE models and obtained consistent improvements, including new state-of-the-art results for the Voice Bank+DEMAND dataset.
\end{abstract}
\section{Introduction}\label{sec:introduction}
Speech Enhancement (SE) is a process of making deteriorated speech signals more understandable and perceptually pleasing. The SE has been widely used for various applications, including mobile communication, speech recognition systems, hearing aids, etc. SE as an area of research interest has been around for several decades. Traditional SE techniques~\cite{1163209, 1163086} often use a heuristic or straightforward signal processing algorithm to estimate a gain function, which is then applied to the noisy input to produce improved speech. Recent developments in deep learning have inspired many Deep Neural Network (DNN)-based SE techniques~\cite{9054563, tran2020single, https://doi.org/10.48550/arxiv.1703.09452, pmlr-v97-fu19b, Wisdom2019DifferentiableCC} that outperform conventional signal processing-based methods. One particular DNN-based architecture, Generative Adversarial Net (GAN), has garnered much interest in the SE community for the past few years~\cite{https://doi.org/10.48550/arxiv.1703.09452, pmlr-v97-fu19b, fu2021metricgan+, cao2022cmgan}. In the applications of SE, GAN architecture is primarily employed to generate enhanced speech. One of the earliest works where GAN models were implemented on the SE domain is the SEGAN~\cite{https://doi.org/10.48550/arxiv.1703.09452} model. It utilizes an adversarial framework to map the noisy waveform to a corresponding enhanced speech. Later, MetricGAN~\cite{pmlr-v97-fu19b} introduced a metric score optimization scheme, where an evaluated metric was introduced into adversarial loss functions, replacing a traditional binary-classifier~\cite{https://doi.org/10.48550/arxiv.1703.09452} and creating a new branch for SE GAN-based research. There have been several improvements to the MetricGAN model, e.g. MetricGAN+~\cite{fu2021metricgan+}, iMetricGAN~\cite{Li2020iMetricGANIE}, CMGAN~\cite{cao2022cmgan}, etc. More recently, with a rise of Transformers~\cite{NIPS2017_3f5ee243} and Conformers ~\cite{Gulati2020ConformerCT}, models such as DB-AIAT~\cite{9746273}, DPT-FSNet~\cite{9746171}, SE-Conformer~\cite{kim21h_interspeech}, CMGAN~\cite{cao2022cmgan}, etc. show significant improvements on SE tasks.

Despite much work, training of GAN-based models are prone to problems such as non-convergence, overfitting, and gradient instabilities. One  common issue in GAN's discriminator training is potentially ``harmful'' gradient direction~\cite{Zadorozhnyy_2021_CVPR} where parts of the model might train opposite to the desired direction. To overcome this problem, we propose a new method called Self-Correcting (SC) Discriminator Optimization. At the same time, the SE DNN-based models are subject to problems caused by the signal-processing tools, e.g., an inconsistency in the Short-Time Fourier Transform (STFT) and its inverse (iSTFT)~\cite{Wisdom2019DifferentiableCC,le2010fast}. Inspired by~\cite{Braun2021TowardsEM}, we adapt and introduce the consistency loss function as a part of Consistency Preserving (CP) Net into the GAN framework, where loss and architecture take into account the iSTFT effects. From our experiments, the combination of SC and CP methods improves the SE GAN-based models even further than either single method; we call such a combination SCP-GAN.


The remainder of this paper is laid out as follows. In section \ref{sec:relatedwork}, we list earlier works pertinent to our work. In section \ref{sec:scpgan}, we introduce our improvements to current GAN-based SE models. We present and compare the SCP-GAN results on Voice Bank+DEMAND dataset~\cite{ValentiniBotinhao2016InvestigatingRS} to the current state-of-the-art (SOTA) models in section \ref{sec:results}. Then, in section \ref{sec:ablstudy}, we provide an extensive ablation study to show the advantages of the proposed methods. Finally, in section \ref{sec:conclusion}, we highlight the methods' contributions to the field.

\section{Related Work}\label{sec:relatedwork}


\subsection{Adaptively Weighted GAN (awGAN)}\label{ss:awgan}
The discriminator plays a very important role in training GAN-based models. However, optimization of the discriminator loss function(s) has been a challenge~\cite{Zadorozhnyy_2021_CVPR}. In the image generation domain, a majority of discriminator loss functions have the following form: $\L_D = \L_r + \L_f$, where $\L_r$ is the part that only relies on the original dataset; \cite{Zadorozhnyy_2021_CVPR} calls it the `real part'. $\L_f$ depends on the generator network, its output, and not the original data; \cite{Zadorozhnyy_2021_CVPR} calls this one the `fake part'. However, the training with $\L_D$ is not performed equally on the real and fake parts, but it depends on the angle between $\nabla \L_r$ and $\nabla \L_f$ and their magnitudes. Under such conditions, the actual training direction $\nabla \L_D$ might end up being in the opposite direction to either $\nabla \L_r$ or $\nabla \L_f$, which is undesirable. To solve such issue~\cite{Zadorozhnyy_2021_CVPR} proposed the method of adaptive weights for the discriminator loss function: $\L_D^{aw} = w_r\L_r + w_f\L_f$, and the algorithm for choosing these weights.

\subsection{STFT Consistencies in SE DNN models}

In audio signal processing, the short-time Fourier transform (STFT) is one of the most fundamental and widely used methods. Most DNN-based SE models~\cite{Wisdom2019DifferentiableCC, cao2022cmgan, le2010fast} use a complex-valued STFTs generator to suppress noise and preserve speech. However, using STFT methods has its issues. One of those issues is the STFT consistency. This is an issue when a loss function does not consider iSTFT signal reconstruction.
Several works have been done to resolve this issue. \cite{le2010fast} presented an algorithm for a phase reconstruction based on a local approximation of the consistency constraints. Adding simple differentiable projection layers to the enhancement DNN to solve the issue was proposed by~\cite{Wisdom2019DifferentiableCC}. More recently, \cite{Braun2021TowardsEM} introduced the iSTFT into back-propagation methods for SE DNN-based models.

\section{SCP-GAN}\label{sec:scpgan}
We propose the following two innovative learning strategies to enhance the performance of the SE GAN-based models. 

\subsection{Self-Correcting Discriminator Optimization}\label{ss:scdo}

We introduce the Self-Correcting (SC) Discriminator Optimization method, an adaptation and generalization of the method from~\cite{Zadorozhnyy_2021_CVPR} to the SE domain. A large number of existing SE GAN-based models have the discriminator loss function consisting
of either two~\cite{pmlr-v97-fu19b, cao2022cmgan} or three~\cite{fu2021metricgan+} equally weighted parts: 

\begin{align}
    \L_D &= \L_C + \L_E \label{eq:2losses}\\
    \L_D &= \L_C + \L_E + \L_N,\label{eq:3losses}
\end{align}
where 
$\L_C$, $\L_E$, and $\L_N$ exclusively rely on clean, enhanced, and noisy datasets, respectively; for example, MetricGAN~\cite{pmlr-v97-fu19b} has a two-part discriminator loss with 

\begin{align}
    \L_C &= \mathbb{E}_y\left(D(y,y)-Q(y,y)\right)^2 \\
    \L_E &= \mathbb{E}_{x,y}\left(D(G(x),y) - Q(G(x),y)\right)^2,
\end{align}
where $x$ is a noisy signal, $y$ is its corresponding clean version, and $D$, $G$, and $Q$ are the discriminative model, generative model, and evaluation metric function, respectively. However, gradient descent training with  $\nabla \L_D$ is not performed equally on clean and enhanced parts; its effect depends on the angle between $\nabla \L_C$ and $\nabla \L_E$ and their magnitudes. For example, if the angle between $\nabla \L_C$ and $\nabla \L_E$ is a large obtuse angle and $\norm{\nabla \L_C}_2 >> \norm{\nabla \L_E}_2$, then $\nabla \L_D$ would make an obtuse angle with $\nabla \L_E$ and thus training along $\nabla \L_D$ would increase the loss $\L_E$, 
which would be harmful to the enhanced part of the model. In addition, MetricGAN+~\cite{fu2021metricgan+} uses the $\L_N$ of the following form
\begin{equation}
    \L_N = \mathbb{E}_{x,y}\left(D(x,y)-Q(x,y)\right)^2.
\end{equation}

To address this issue, as in~\cite{Zadorozhnyy_2021_CVPR} for GAN, we introduce weights into the two-part discriminator in (\ref{eq:2losses}),
\begin{equation}
    \L_D^{SC} = w_C\L_C + w_E\L_E \label{eq:2losses_sc}
\end{equation}
and call it SC$_2$. Moreover, we
introduce weights into the three-part discriminator in (\ref{eq:3losses})
\begin{equation}
    \L_D^{SC} = w_C\L_C + w_E\L_E + w_N\L_N \label{eq:3losses_sc}
\end{equation}
and call it SC$_3$. The weights in (\ref{eq:2losses_sc}) and (\ref{eq:3losses_sc}) are determined by Algorithm \ref{alg:sc} using $\nabla\L_C$, $\nabla\L_E$, or $\nabla\L_N$, resulting in a self-correcting discriminator gradient. The goal of Algorithm \ref{alg:sc} is to minimize the potential harm to parts of the model by ensuring the training gradient $\nabla \L_D^{SC}$ does not go obtuse to $\nabla\L_C$, $\nabla\L_E$, or $\nabla\L_N$.

\vspace{-0.1in}
\IncMargin{0.0em}\SetNlSty{text}{}{:}
\begin{algorithm}
    \scriptsize
    \SetAlgoLined
        \textbf{Compute:} $\nabla \L_C$, $\nabla \L_E$, $\nabla \L_N$ \textbf{if} $\L_N$ is used 
        
        \eIf{$\angle_2\left(\nabla \L_C, \nabla \L_E\right) < 90^{\circ}$}
        {
        $w_C = 1$ and $w_E = 1$
        
            \eIf{$\L_N$ is used \textnormal{\textbf{and}} \resizebox{0.65\hsize}{!}{$\angle_2\left(w_C\nabla \L_C+ w_E\nabla \L_E, \nabla \L_N\right)< 90^{\circ}$}}
            {
            $w_N = 1$
            }{
            $w_N = -\dfrac{\la \nabla \L_C, \nabla \L_N \ra_2}{\norm{\nabla \L_N}_2^2}-\dfrac{\la\nabla \L_E, \nabla \L_N \ra_2}{\norm{\nabla \L_N}_2^2}$
            }
        }{
        $w_C = 1$ and $w_E = -\dfrac{\la \nabla \L_C, \nabla \L_E \ra_2}{\norm{\nabla \L_E}_2^2}$

            \eIf{$\L_N$ is used \textnormal{\textbf{and}} \resizebox{0.65\hsize}{!}{$\angle_2\left(w_C\nabla \L_C+ w_E\nabla \L_E, \nabla \L_N\right)< 90^{\circ}$}}
            {
            $w_N = 1$
            }{
            \resizebox{1.05\hsize}{!}{
            $w_N = -\dfrac{\la \nabla \L_C, \nabla \L_N \ra_2}{\norm{\nabla \L_N}_2^2} + \dfrac{\la\nabla\L_C,\nabla\L_E \ra_2\la\nabla\L_E,\nabla\L_N \ra_2}{\norm{\nabla \L_E}_2^2\norm{\nabla \L_N}_2^2}$}
            }
        }
    \caption{SC$_2$/SC$_3$ method}
    \label{alg:sc}
\end{algorithm}

\subsection{Consistency Preserving Network}\label{ss:cpgen}

\begin{figure*}[ht!]
    \centering
    \subfloat[Process of computing Time and TF-magnitude losses inside the GAN-based SE model Generator (G)]{
    \raisebox{0.2in}{\includegraphics[width=0.95\columnwidth]{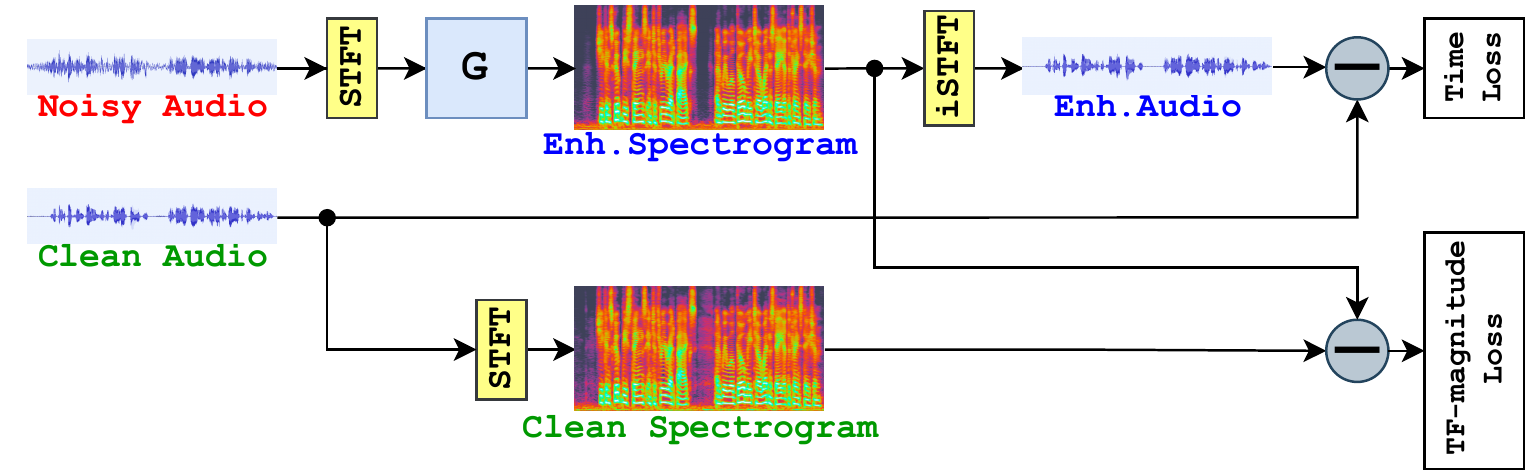}}\label{fig:not_cpn}}
    \hfill
    \subfloat[\textbf{Consistency Preserving (CP) Net:} Depiction of the process for computing Time and TF-magnitude losses with CP method inside the GAN-based SE model Generator (G)]{
    {\includegraphics[width=0.95\columnwidth]{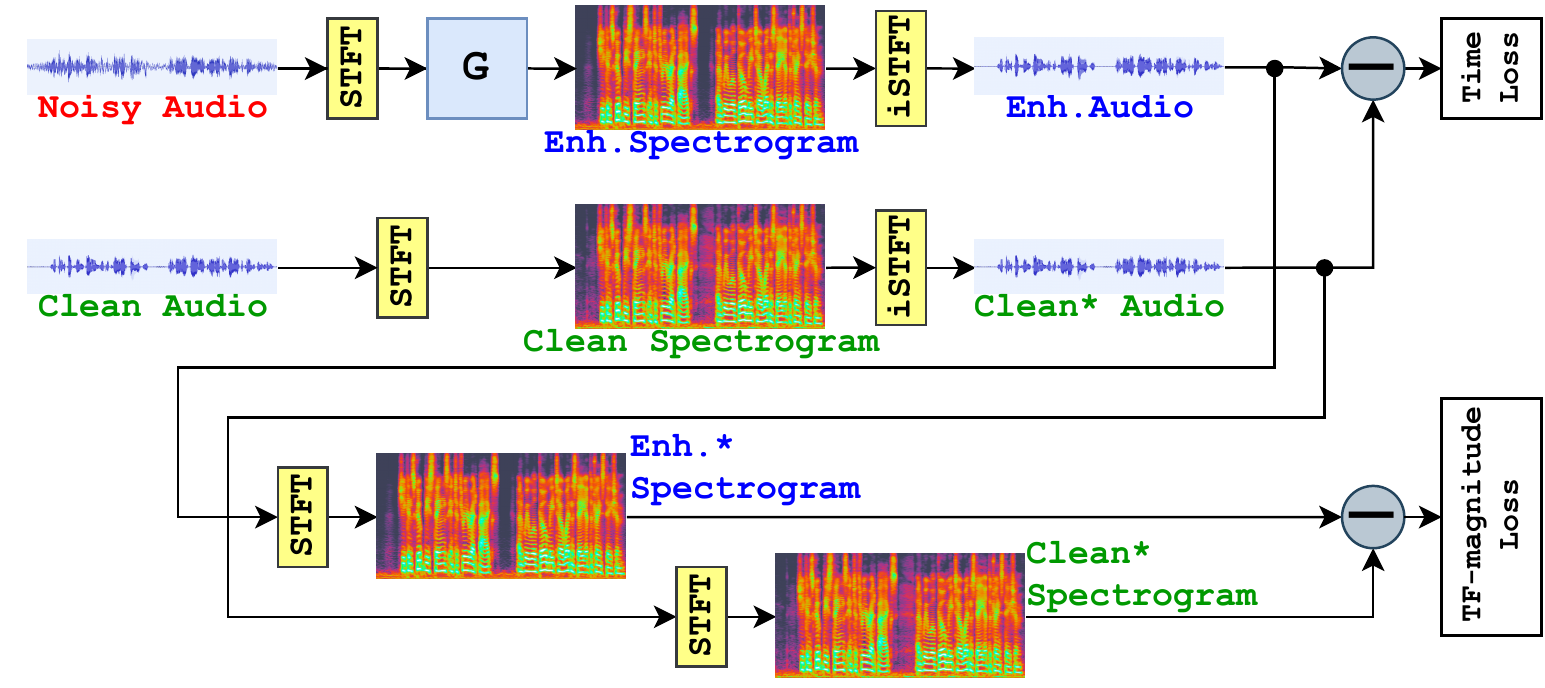} }\label{fig:cpn}}
    \label{fig:cpn_total}
    \caption{Traditional vs. Consistency Preserving SE GAN-based models}
\end{figure*}

The majority of GAN-based SE models~\cite{pmlr-v97-fu19b, cao2022cmgan, 9746273, 9746171} have a generator (G) that accepts STFT spectrogram of a noisy waveform as input. The G's output is an enhanced spectrogram that later uses iSTFT to produce the enhanced waveform; Figure \ref{fig:not_cpn} illustrates the process. The G is then updated using a combination of various loss functions, e.g. Time Loss~\cite{cross_domain_losses_for_se}, TF-magnitude Loss~\cite{Braun2021ACV}, Adversarial Metric Loss~\cite{pmlr-v97-fu19b}, etc. For example, the TF-magnitude Loss~\cite{Braun2021ACV} is computed between the enhanced and clean spectrograms, see diagram in Figure \ref{fig:not_cpn}. However, such loss and architectural setup do not take into account the effect of the iSTFT reconstruction 
 which causes inconsistencies between signals. 


We incorporate the idea from~\cite{Braun2021TowardsEM} to SE GAN-based model by modifying architecture and loss function(s) such that any input into a loss function undergoes the same process, taking into consideration the effects of signal reconstruction from the spectrogram; we call such process and loss function a Consistency Preserving (CP) Network and a consistency loss, respectively. Figure \ref{fig:cpn} depicts the process of computing Time and TF-magnitude losses using the proposed CP method by ensuring that the same transforms are applied on enhanced, clean, and noisy (if used) signals. 

Note that the Clean Audio to the Clean$^*$ Audio process inside the CP Net (2$^{\text{nd}}$ row of Figure \ref{fig:cpn}) can be performed at the data preprocessing stage.

\section{Experiments}\label{sec:results}

\begin{table*}[t]
    \centering
    \resizebox{0.775\textwidth}{!}{%
    \begin{tabular}{l||c|c|c|c|c|c|c}
        \toprule
        Model & \# of Param. & PESQ & CSIG & CBAK & COVL & SSNR & STOI\\
        \midrule
        Noisy Data & n/a & 1.97 & 3.35 & 2.44 & 2.63 & 1.68 & 0.91 \\
        \midrule
        SE-Conformer~\cite{kim21h_interspeech} & - & 3.13 & 4.45 & 3.55 & 3.82 & - & 0.95 \\
        MANNER~\cite{park2022manner} & - & 3.21 & 4.53 & 3.65 & 3.91 & - & - \\
        DB-AIAT~\cite{9746273} & 2.81M & 3.31 & 4.61 & 3.75 & 3.96 & 10.79 & 0.96 \\
        DPT-FSNet~\cite{9746171} & 0.91M & 3.33 & 4.58 & 3.72 & 4.00 & - & 0.96 \\
        PCS~\cite{chao2022perceptual} & - & 3.35 & 4.43 & - & 3.92 & - & 0.95 \\
        \midrule
        MetricGAN+~\cite{fu2021metricgan+} & 2.6M & 3.15 & 4.14 & 3.16 & 3.64 & - & 0.93$^\dagger$ \\
        MetricGAN+ (repro.) & 2.6M & 3.08 & 4.05 & 3.01 & 3.60 & - & 0.92 \\
        SCP-MetricGAN+ (ours) & 2.6M & 3.19 & 4.20 & 3.20 & 3.65 & - & 0.93 \\
        \midrule
        CMGAN~\cite{cao2022cmgan} &  1.83M & 3.41 & 4.63 & 3.94 & 4.12 & \bf{11.10} & 0.96\\
        CMGAN (repro.) & 1.83M & 3.39 & 4.62 & 3.93 & 4.13 & 10.61 & 0.96\\
        SCP-CMGAN (ours) & 1.83M & \bf{3.52} & \bf{4.75} & \bf{3.97} & \bf{4.25} & 10.82 & \bf{0.96}\\
        \bottomrule
    \end{tabular}
    }
    \caption{\textbf{Performance comparison on Voice Bank+DEMAND dataset~\cite{ValentiniBotinhao2016InvestigatingRS}:} ``-'' denotes the results not provided in the original paper; $\dagger$ - quoted from~\cite{chao2022perceptual}; repro. - our reproduction of experiments}
    \label{tab:vbd_results}
\end{table*}

\subsection{Dataset}
We use the publicly accessible Voice Bank+DEMAND~\cite{ValentiniBotinhao2016InvestigatingRS} dataset to evaluate and compare our proposed SCP-GAN method. The training set of the Voice Bank+DEMAND dataset consists of 11,572 individual recordings of 28 speakers from the Voice Bank corpus~\cite{6709856} which are mixed with DEMAND~\cite{doi:10.1121/1.4799597} database and some artificial background noises at the signal-to-noise ratios (SNRs) of 0, 5, 10, and 15 dB. The test set has 824 utterances of two speakers from the Voice Bank corpus, which are mixed with unseen DEMAND noises at the SNRs of 2.5, 7.5, 12.5, and 17.5 dB.

\subsection{Evaluation Metrics}
To assess the speech quality, we select a set of widely used metrics, including the Perceptual Evaluation of Speech Quality (PESQ)~\cite{Rix2001PerceptualEO} (ranging between -0.5 and 4.5), the Mean Opinion Score (MOS)~\cite{4389058}, prediction of the signal distortion (CSIG), the MOS prediction of the intrusiveness of background noise (CBAK), MOS prediction of the overall effect (COVL) (all MOS metrics range between 1 and 5), the Segmental Signal-to-Noise Ratio (SSNR), and the Short-Time Objective Intelligibility (STOI)~\cite{5495701} (with a range 0 to 1). For all metrics, higher numbers denote better performance.

\subsection{Experimental Results}
We have applied our proposed methods to two baseline models: a widely-used MetricGAN+~\cite{fu2021metricgan+} model and a current SOTA model - CMGAN~\cite{cao2022cmgan}. Our SCP method 
shows a consistent improvement over the compared baseline. On the MetricGAN+ model, our SCP-MetricGAN+ improved by 0.04, 0.06, 0.04, and 0.01 on the PESQ, CSIG, CBAK, and COVL metrics, respectively. Our improvements with the SCP-CMGAN model are 0.11, 0.12, 0.03, and 0.13 on the same scores. Moreover, we have compared our method to other recent SOTA models which can be seen in Table \ref{tab:vbd_results}.

\textbf{Note:} Results provided in Table \ref{tab:vbd_results} for MetricGAN+ and CMGAN models are quoted from the original papers; however, to verify them, we have obtained our results: MetricGAN+ (repro.) and CMGAN (repro.). The results for the CMGAN (repro.) model are very similar to the results from~\cite{cao2022cmgan} with one exception - the SSNR metric, where our result is lower: 10.61 (ours) vs. 11.10~\cite{cao2022cmgan}. Furthermore, the results for MetricGAN+ (repro.) model are slightly lower than the ones provided in the original paper~\cite{fu2021metricgan+}.

\section{Ablation Study}\label{sec:ablstudy}

We have conducted an ablation study to demonstrate the importance of our methods. We have chosen the CMGAN~\cite{cao2022cmgan} model as the base model due to its SOTA performance at the time of this study. Table \ref{tab:abs_study} shows the average results of each model's best performance over three randomly chosen seeds.

First, we have retrained the CMGAN~\cite{cao2022cmgan} model to verify the results from~\cite{cao2022cmgan}. The results obtained from our experiments are relatively close to the results stated in~\cite{cao2022cmgan}, except for the SSNR metric where we have obtained slightly lower results, i.e., SSNR of 10.61 (ours) vs. SSNR of 11.10~\cite{cao2022cmgan}.

Next, we have added Noisy Data (ND) to the CMGAN model discriminator training (`+ ND' in Table \ref{tab:abs_study}); however, such an addition had slight improvement over the baseline. Following it, we have considered adding our SC method 
to the baseline model (`+ SC$_{2}$' in Table \ref{tab:abs_study}) and nothing else. 
With SC$_{2}$, we saw some improvements in PESQ, COVL, and SSNR metrics. Furthermore, we have analyzed the advantages of the CP method (`+ CP' in Table \ref{tab:abs_study}) without any add-ons. The CP shows significant improvements in PESQ, CSIG, and COVL metrics and is comparable in the others.

Then, we combined ND and SC$_3$ methods (`+ ND, SC$_3$' in Table \ref{tab:abs_study}).
Such a setup further improves baseline as well as single methods, particularly in COVL and SSNR metrics. A combination of ND and CP methods (`+ ND, CP' in Table \ref{tab:abs_study}) has the same nature of improvements, producing better results in PESQ, CSIG, and COVL metrics. The last combination of SC$_2$ and CP methods (`+ SC$_2$, CP' in Table \ref{tab:abs_study}) demonstrates that together both proposed methods achieve significant improvements on the SE task. Moreover, this particular model achieved the highest SSNR result of 10.91.

Finally, we have combined ND, SC$_3$, and CP methods in the model we call SCP-CMGAN in Table \ref{tab:vbd_results}. Adding ND and switching from SC$_2$ to SC$_3$ further improves the `CMGAN + SC$_2$, CP' model, achieving new SOTA results.

Note that all of the above models were trained under the same conditions without changing the hyperparameters and with identical software and hardware settings: Python 3.8.13, PyTorch 1.10, and CUDA 11 on NVIDIA Tesla V100 GPUs.

\begin{table}[t]
    \centering
    \resizebox{\columnwidth}{!}{%
    \begin{tabular}{l|c|c|c|c|c}
        \toprule
        Model & PESQ & CSIG & CBAK & COVL & SSNR\\
        \midrule
        CMGAN (repro.)$^\dagger$ & 3.39 & 4.62 & 3.93 & 4.13 & 10.61 \\
        \midrule
        \, + ND & 3.41 & 4.65 & 3.92 & 4.13 & 10.68 \\
        \, + SC$_2$ & 3.44 & 4.65 & 3.92 & 4.17 & 10.70 \\
        \, + CP & 3.47 & 4.71 & 3.93 & 4.20 & 10.54 \\
        \midrule
        \, + ND, SC$_3$ & 3.43 & 4.64 & 3.93 & 4.18 & 10.76 \\
        \, + ND, CP & 3.47 & 4.73 & 3.93 & 4.22 & 10.53 \\
        \, + SC$_2$, CP & 3.49 & 4.72 & 3.96 & 4.24 & \textbf{10.91} \\
        \midrule
        \, + ND, SC$_3$, CP & \textbf{3.52} & \textbf{4.75} & \textbf{3.97} & \textbf{4.25} & 10.82 \\
        \bottomrule
    \end{tabular}
    }
    \caption{\textbf{Ablation Study on Voice Bank + DEMAND:} STOI results are equal to 0.96 for all the tests; $^\dagger$ - results from our tests; ND - Noisy Data, CP - Consistency Preserving method, SC$_{2}$ - SC with $\L_C$ and $\L_E$, SC$_{3}$ - SC with $\L_C$, $\L_E$, $\L_N$.}
    \label{tab:abs_study}
\end{table}

\section{Conclusion}\label{sec:conclusion}
This paper presents several improvements to SE GAN-based models. The proposed method of Consistency Preservation reconciles the issue with Fourier and Inverse-Fourier transforms inside the generative models. At the same time, the Self-Correcting Discriminator Optimization method helps with training the discriminative model by avoiding gradient directions that are potentially harmful to the training. Our experiments and ablation study demonstrate the advantages of using one or both of the proposed methods, including new SOTA results for the Voice Bank+DEMAND dataset.

\clearpage
\bibliographystyle{IEEEbib}
\bibliography{refs_short.bib}

\end{document}